\documentclass{article}

\usepackage{amsmath,amsfonts,amssymb}
\usepackage{graphicx}
\usepackage{color}
\usepackage{subfigure}

\newcommand{\re}{{\mathbb{R}}}
\newcommand{\V}{V}
\newcommand{\ipr}[2]{\langle #1, #2 \rangle}
\newcommand{\vc}[1]{{\boldsymbol{#1}}}
\newcommand{\elll}{\ell^1\text{-}\ell^2}
\newcommand{\jj}{{\mathrm{j}}}
\newcommand{\dd}{{\mathrm{d}}}
\newcommand{\xc}{\vc{x}_{\text{c}}}
\newcommand{\dxc}{\dot{\vc{x}}_{\text{c}}}
\newcommand{\triplenorm}{\ensuremath{| \! | \! |}}
\newcommand{\tnorm}[1]{\triplenorm #1 \triplenorm}
\newcommand{\rms}[1]{\mathrm{RMS}(#1)}

\newtheorem{prob}{Problem}

\begin{document}
\title{\huge \bf Compressive Sampling for\\Networked Feedback Control}
\author{Masaaki Nagahara%
\thanks{
Graduate School of Informatics, Kyoto University,
{\tt nagahara@ieee.org}
(corresponding author)
},~
Daniel E. Quevedo%
\thanks{
School of Electrical Engineering \& Computer Science,
The University of Newcastle
},\\
Takahiro Matsuda%
\thanks{
Graduate School of Engineering, Osaka University
},~
Kazunori Hayashi%
\thanks{
Graduate School of Informatics, Kyoto University
}
}
\maketitle
\begin{abstract}
We investigate the use  of compressive sampling 
for networked feedback control systems.
The method proposed serves to compress the control vectors
which are transmitted through rate-limited channels
without much deterioration of control performance.
The control vectors are obtained by an $\elll$ optimization,
which can be solved very efficiently by FISTA 
(Fast Iterative Shrinkage-Thresholding Algorithm).
Simulation results show that the proposed sparsity-promoting 
control scheme gives a better control performance
than a conventional energy-limiting $L^2$-optimal control.
\end{abstract}

\section{Introduction}
The objective of this article is
to design a controller in a {\em networked control system}
\cite{BemHeeJoh}
that produces sparse control vectors for effective compression
before transmissions.
Unfortunately, the calculation of optimal sparse vectors will, in general, require
significant computational cost and may thereby introduce delays, 
which are unacceptable for closed-loop operation. 
To overcome this issue, we subsample the problem
to reduce its size and adopt a fast algorithm
called \emph{FISTA} (Fast Iterative Shrinkage-Thresholding Algorithm) \cite{BecTeb09}.

Networked control systems 
are those
in which the controlled plants are located away from
the controllers, and the communication should be made
through rate-limited communication channels such as wireless
networks or the Internet \cite{Tan}.
In networked control systems, efficient signal compression
or representation is essential to send control data
through rate-limited communication channels.
For this purpose, we propose an approach of sparse control signal 
representation using the {\em compressive sampling} technique 
\cite{Can06}.
Our contributions in this paper are
(1) a new strategy for networked feedback control systems
based on compressive sampling,
(2) an effective data compression scheme of the control signals
with sparse representation,
(3) formulation of the design problem by $\elll$ optimization which can be
efficiently solved.

The compressive sampling approach will open up a new vista in control theory.
To the best of our knowledge, 
so far only a few studies have applied compressive sampling to control:
\cite{BhaBas11} proposes to use compressive sensing in feedback control systems
for perfect state estimation and
\cite{NagQue11} proposes sparse representation of transmitted control packets
for feedback control with packet dropouts.
For remote control systems, \cite{NagQueOstMatHay11,NagMatHay12} also propose
to use $\elll$ optimization (as in this paper).
However, \cite{NagQueOstMatHay11,NagMatHay12} consider only feed-forward control systems.

\section{Control Problem}
\label{sec:ncs}
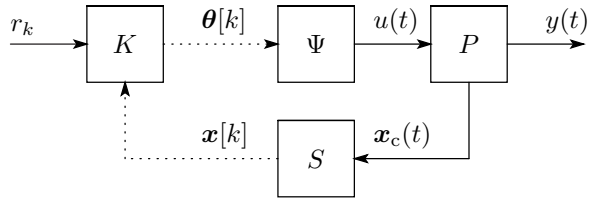
\begin{figure}[tb]
\centering{
\unitlength 0.1in
\begin{picture}( 30.0000, 10.0000)(  4.0000,-12.0000)
%
\special{pn 8}%
\special{pa 400 400}%
\special{pa 800 400}%
\special{fp}%
\special{sh 1}%
\special{pa 800 400}%
\special{pa 734 380}%
\special{pa 748 400}%
\special{pa 734 420}%
\special{pa 800 400}%
\special{fp}%
%
\special{pn 8}%
\special{pa 800 200}%
\special{pa 1200 200}%
\special{pa 1200 600}%
\special{pa 800 600}%
\special{pa 800 200}%
\special{pa 1200 200}%
\special{fp}%
%
\special{pn 8}%
\special{pa 1200 400}%
\special{pa 1800 400}%
\special{dt 0.045}%
\special{sh 1}%
\special{pa 1800 400}%
\special{pa 1734 380}%
\special{pa 1748 400}%
\special{pa 1734 420}%
\special{pa 1800 400}%
\special{fp}%
%
\special{pn 8}%
\special{pa 1800 200}%
\special{pa 2200 200}%
\special{pa 2200 600}%
\special{pa 1800 600}%
\special{pa 1800 200}%
\special{pa 2200 200}%
\special{fp}%
%
\special{pn 8}%
\special{pa 2200 400}%
\special{pa 2600 400}%
\special{fp}%
\special{sh 1}%
\special{pa 2600 400}%
\special{pa 2534 380}%
\special{pa 2548 400}%
\special{pa 2534 420}%
\special{pa 2600 400}%
\special{fp}%
%
\special{pn 8}%
\special{pa 2600 200}%
\special{pa 3000 200}%
\special{pa 3000 600}%
\special{pa 2600 600}%
\special{pa 2600 200}%
\special{pa 3000 200}%
\special{fp}%
%
\special{pn 8}%
\special{pa 3000 400}%
\special{pa 3400 400}%
\special{fp}%
\special{sh 1}%
\special{pa 3400 400}%
\special{pa 3334 380}%
\special{pa 3348 400}%
\special{pa 3334 420}%
\special{pa 3400 400}%
\special{fp}%
%
\special{pn 8}%
\special{pa 2800 600}%
\special{pa 2800 1000}%
\special{fp}%
%
\special{pn 8}%
\special{pa 2800 1000}%
\special{pa 2200 1000}%
\special{fp}%
\special{sh 1}%
\special{pa 2200 1000}%
\special{pa 2268 1020}%
\special{pa 2254 1000}%
\special{pa 2268 980}%
\special{pa 2200 1000}%
\special{fp}%
%
\special{pn 8}%
\special{pa 2200 800}%
\special{pa 1800 800}%
\special{pa 1800 1200}%
\special{pa 2200 1200}%
\special{pa 2200 800}%
\special{pa 1800 800}%
\special{fp}%
%
\special{pn 8}%
\special{pa 1800 1000}%
\special{pa 1000 1000}%
\special{dt 0.045}%
%
\special{pn 8}%
\special{pa 1000 1000}%
\special{pa 1000 600}%
\special{dt 0.045}%
\special{sh 1}%
\special{pa 1000 600}%
\special{pa 980 668}%
\special{pa 1000 654}%
\special{pa 1020 668}%
\special{pa 1000 600}%
\special{fp}%
\put(10.0000,-4.0000){\makebox(0,0){$K$}}%
\put(20.0000,-4.0000){\makebox(0,0){$\Psi$}}%
\put(28.0000,-4.0000){\makebox(0,0){$P$}}%
\put(20.0000,-10.0000){\makebox(0,0){$S$}}%
\put(4.0000,-3.5000){\makebox(0,0)[lb]{$r_k$}}%
\put(14.0000,-3.5000){\makebox(0,0)[lb]{$\vc{\theta}[k]$}}%
\put(14.0000,-9.5000){\makebox(0,0)[lb]{$\vc{x}[k]$}}%
\put(23.0000,-9.5000){\makebox(0,0)[lb]{$\xc(t)$}}%
\put(23.0000,-3.5000){\makebox(0,0)[lb]{$u(t)$}}%
\put(32.0000,-3.5000){\makebox(0,0)[lb]{$y(t)$}}%
\end{picture}%
}
\caption{Networked control system. The dotted line indicates a rate-limited communication channel.}
\label{fig:ncs}
\end{figure}

Fig.~\ref{fig:ncs} shows the networked control system which we consider 
in this article.
The system consists of a controlled plant $P$,
a sensor (or sampler) $S$, a decoder or a digital-to-analog (DA) converter $\Psi$,
and a digital controller $K$.
The definitions of these systems are given as follows:
\begin{description}
\item[Plant $P$: ]
The controlled plant $P$ is modeled
by the following state-space representation:
\begin{equation}
 P:\left\{ \begin{split}
	    \dxc(t) &= A\xc(t) + \vc{b}u(t),\\ 
	   y(t) &= \vc{c}^\top \xc(t),~t\in[0,\infty),
 \end{split}\right.
 \label{eq:plant}
\end{equation}
where $\xc(0)=\vc{0}$,
$A\in\re^{\nu\times \nu}$, and $\vc{b},\vc{c} \in\re^{\nu\times 1}$.
\item[Sensor $S$: ]
The sensor (or sampler) $S$ converts the continuous-time
state $\xc$ into a discrete-time signal $\vc{x}[k]:=\xc(kT)$,
$k=0,1,2,\dots$, where $T>0$
is the sampling period.
\item[Decoder $\Psi$: ]
The decoder (or DA converter) $\Psi$ converts
a vector valued signal $\vc{\theta}[k]$ into a continuous-time signal $\{u_k(t)\}_{t\in[0,T)}$,
$k=0,1,2,\dots$
via
\begin{equation}
  \Psi: \vc{\theta}[k] \mapsto u_k := \sum_{m=-M}^M \theta_m[k]\psi_m \in L^2[0,T),
 \label{eq:Psi}
\end{equation}
where $M$ is a positive integer, $\theta_m[k]$ is the $m$-th element of
the vector $\vc{\theta}[k]$,
and 
\begin{equation}
 \psi_m(t) := \frac{1}{\sqrt{T}}\exp({\jj\omega_m t}),~\omega_m := \frac{2\pi m}{T},~t\in[0,T).
 \label{eq:exp}
\end{equation}
We call the vector $\vc{\theta}[k]$ a {\em control vector}.
Note that the continuous-time signal $u_k$ is band-limited to $\omega_M=2\pi M/T$ [rad/sec].
That is, $u_k$ belongs to the following subspace of $L^2[0,T)$:
\begin{equation}
 \V_M := \mathrm{span}\{\psi_{-M},\dots,\psi_{M}\} \subset L^2[0,T].
 \label{eq:VM}
\end{equation}
The input $u$ to the plant $P$ is defined by
$u(t+kT) = u_k(t)$, $t\in[0,T)$, $k=0,1,2,\dots$.
\item[Controller $K$: ]
The controller $K$ uses
a continuous-time reference signal
$r_k\in V_M$, $k=0,1,2,\dots$,
and the sampled state $\vc{x}[k]$ to
produce
the control vector $\vc{\theta}[k]$.
The latter defines the input signal $u_k$ as per (\ref{eq:Psi}).
\end{description}

We assume that $S$, $\Psi$, and $K$ are synchronized at $t=kT$, $k=0,1,2,\dots$.
We also assume that we can transmit the control vector 
$\vc{\theta}[k]$ and the sampled state $\vc{x}[k]$ through communication channels
without any delays nor packet dropouts.
In this article, we consider a situation 
where the size $N=2M+1$ of $\vc{\theta}[k]$ is much larger than the size $\nu$ of
the state $\vc{x}[k]$, and should be compressed because $\vc{\theta}[k]$ 
needs to be transmitted through a rate-limited communication channel.

Under these assumptions, we then formulate our control problem.
Let $y_k$ be the continuous-time signal $y$ on the interval $[kT, (k+1)T)$,
that is, $y_k(t):=y(t+kT)$, $t\in[0,T)$, $k=0,1,\dots$.
We design the controller $K$ to achieve the following objectives:
\begin{enumerate}
\item The first objective is to attenuate the tracking error
between the reference $r_k$ and the output $y_k$ on the interval $[kT, (k+1)T)$,
$k=0,1,2,\dots$.
The error is measured by the $L^2$ norm:
\[
 \tnorm{y_k-r_k}_2^2 := \int_0^T \left|y_k(t)-r_k(t)\right|^2 \dd t.
\]
\item The second objective is to reduce the data size
of the control vector $\vc{\theta}[k]$ which defines the control $u_k$ via (\ref{eq:Psi}).
For this objective, we adopt the so-called $0$-norm of $u_k$ defined by
$\tnorm{u_k}_0 := \|\vc{\theta}[k]\|_0$, 
the number of the nonzero elements in $\vc{\theta}[k]$.
\end{enumerate}
In general, there is a trade-off between tracking-error attenuation
and data-size reduction.
For example, the sparsest solution $u_k\equiv 0$ leads to very large error,
and the control $u_k$ which minimizes only the first objective function may not be sparse.
To solve this problem, we adopt {\em regularization}.
The problem is formulated as follows.
\begin{prob}
\label{prob1}
{\rm
Given reference signal $r_k\in\V_M$, $k=0,1,2,\dots$,
find the control $u_k\in\V_M$ (or the control vector $\vc{\theta}[k]$) which minimizes
\begin{equation}
J(u_k):=\tnorm{y_k-r_k}_2^2 + \mu \tnorm{u_k}_0,
 \label{eq:prob1}
\end{equation}
where $\mu>0$ is the regularization parameter to reconcile the trade-off
between the tracking error and the sparsity.
}
\end{prob}

\section{Compressive sampling for sparse control vectors}
\label{sec:CS}
The objective function $J(u_k)$ in (\ref{eq:prob1}) is defined on an infinite-dimensional
signal subspace $\V_M$ defined in (\ref{eq:VM}).
We here relax the objective function into a finite-dimensional convex $\elll$
optimization problem by using the technique of compressive sampling.

Since the signals $r_k$ and $u_k$ are assumed to be band-limited up to
the frequency $\omega_M=2\pi M/T$ [rad/sec],
we can safely discretize the signals by sampling them at a sampling frequency
higher than $2\omega_M$ based on Shannon's sampling theorem
\cite{Uns00}.
However, if $M$ is very large, it may take very long time to compute the optimal vector.
It follows that there may exist a large delay in the feedback loop,
which may lead to instability and control performance deterioration.
Hence it is preferable to use a more efficient method than Shannon's sampling.
For this purpose, we adopt the technique of compressive sampling
\cite{Can06}
with {\em random sampling},
which can reduce the computational load for the optimization.

Random sampling is modeled as follows:
we first split the interval $[0,T)$ with sampling points
$t_n:=(n-1)/f_M$, $n=1,2,\dots,N=2M+1$,
where $f_M:=2M/T$ is the Nyquist rate.
Then we randomly choose $K$ sampling points ($K<N$) from $\{t_1,\dots,t_N\}$.
To model this, we define a random matrix
$U := [\vc{e}_{i(1)},\vc{e}_{i(2)},\dots,\vc{e}_{i(K)}]^\top  \in \{0,1\}^{K\times N}$,
where $i(1),\dots,i(N)$ are discrete random variables
chosen from the uniform distribution on $\{1,2,\dots,N\}$
such that $i(l)<i(l+1)$, $l=0,1,\dots,N-1$,
and $\{\vc{e}_1,\dots,\vc{e}_N\}$ is the standard basis in $\re^N$, that is,
$\vc{e}_n$ ($n = 1, 2 \dots, N$) denotes a unit vector 
whose $n$-th element is equal to one and the other elements are equal to zero.
The random matrix $U$ is re-chosen at every sampling step $k$.

By using the random variables $i(1), i(2),\dots,i(K)$, we define the random sampling points by
$t_{i(l)} := i(l)h$, $h:=T/(N-1)$, $l=1,2,\dots,K<N$.
Then we consider random sampling of the output $y_k$.
The sampled output $y_k(t_n)$ with the control signal $u_k\in\V_M$ defined in (\ref{eq:Psi})
is computed by
\begin{equation}
 y_k(t_n) = \vc{c}^\top \exp(t_nA)\vc{x}_0 + \sum_{m=-M}^M \theta_m[k] \ipr{\phi_n}{\psi_m},
 \label{eq:ytn}
\end{equation}
where $\ipr{\cdot}{\cdot}$ is the inner product in $L^2[0,T)$, and
\[
  \phi_n(t) := \begin{cases}\vc{c}^\top \exp\left[{(t_n-t)A}\right]\vc{b},\quad &t\in[0,t_n),\\ 0,\quad &t\in(t_n,T].\end{cases}
\]

Define the randomly sampled output vector 
\[
 \vc{y}[k]:=[y_k(t_{i(1)}),\dots,y_k(t_{i(K)})]^\top\in \re^{K}.
\]
Then by (\ref{eq:ytn}), we have
$\vc{y}[k] = UG\vc{\theta}[k] + UH\vc{x}[k]$,
where $G$ is an $N\times N$ matrix defined by 
$(G)_{ij}=\ipr{\phi_i}{\psi_j}$,
$i=1,\dots,N$, $j=-M,\dots,M$,
and $H$ is an $N\times \nu$ matrix defined by
$H := \left[\exp(t_1A^\top)\vc{c},\dots,\exp(t_NA^\top)\vc{c}\right]^\top$.
Let
$\vc{r}[k] := \left[r_k(t_1),r_k(t_2),\dots,r_k(t_N)\right]^\top\in\re^N$
and $\Phi:=UG$, $\vc{\alpha}[k]=U(\vc{r}[k]-H\vc{x}[k])$.
Then the tracking error at the random sampling points
$\{t_{i(1)},t_{i(2)},\dots,t_{i(K)}\}$ is given by
$\vc{y}[k]-\vc{r}[k] = \Phi\vc{\theta}[k]-\vc{\alpha}[k]$.
It follows that the cost function (\ref{eq:prob1}) in Problem \ref{prob1}
is approximately described in a finite-dimensional one:
\begin{equation}
 J_0(\vc{\theta}[k]) := \|\Phi\vc{\theta}[k]-\vc{\alpha}[k]\|_2^2 + \mu \|\vc{\theta}[k]\|_0.
 \label{eq:J0}
\end{equation}

The minimization of the cost function (\ref{eq:J0}) is still difficult to solve
when $M$ is large since the optimization is a combinatorial one.
To reduce this, we adopt a convex relaxation by replacing the $\ell^0$ norm
with the $\ell^1$ norm:
\begin{equation}
 J_1(\vc{\theta}[k]) := \|\Phi\vc{\theta}[k]-\vc{\alpha}[k]\|_2^2 + \mu \|\vc{\theta}[k]\|_1.
 \label{eq:J1}
\end{equation}
The cost function $J_1(\vc{\theta}[k])$ in (\ref{eq:J1}) is convex in $\vc{\theta}[k]$ and hence
the optimal value uniquely exists.
To obtain the $\elll$ optimal vector,
we use an iterative algorithm called
FISTA \cite{BecTeb09}.
This algorithm is very simple and fast;
it can be effectively implemented in digital devices,
which leads to a real-time computation in the feedback loop.
For this algorithm, see \cite{BecTeb09}.

\section{Simulation results}
In this section, we illustrate simulation results to show the effectiveness
of the compressive sampling technique in networked feedback control systems.

The matrices in the state-space representation (\ref{eq:plant}) of the controlled plant $P$
are taken as
\[
 A = \begin{bmatrix}0&1\\-\alpha\beta&-\alpha-\beta\end{bmatrix},~~
 \vc{b}=\begin{bmatrix}0\\1\end{bmatrix},~~
 \vc{c}=\begin{bmatrix}-\alpha\\1\end{bmatrix},
\]
where $\alpha=5$ and $\beta=10$.
We assume the initial state $\vc{x}(0)=\vc{0}$.
The control period $T$ is set to be $2\pi$.
The number of the basis functions $\{\psi_m\}$, or the size of the control vector $\vc{\theta}[k]$ is
$N=2M+1=101$ ($M=50$).
We use the reference
\[
 r_k(t) = \sin(10t) + \cos(5t),~k=0,1,2,\dots.
\]
The sparsity of the reference $r_k$ is given by 
$\tnorm{r_k}_0 = 8 \ll N=101$. 
Therefore, $r_k$ is a sparse vector when it is represented 
by the basis functions $\{\psi_m \}$ defined in (\ref{eq:exp}).
That is, the reference $r_k$ is sparse with respect to the basis $\{\psi_m\}$.
The shortest sampling interval in random sampling is
$h=T/(N-1)=2\pi/100$.
We set the number of random sampling $K=33$.
The iteration steps in FISTA for minimizing the $\elll$ optimization in (\ref{eq:J1})
is 10.
We run the simulation of the feedback control for $k=0,1,\dots,100$,
that is, the length of simulation time is $T_{\text{f}}:=T\times 101 = 202\pi$.

\begin{figure}[tb]
\centering{
\includegraphics[width=0.88\linewidth]{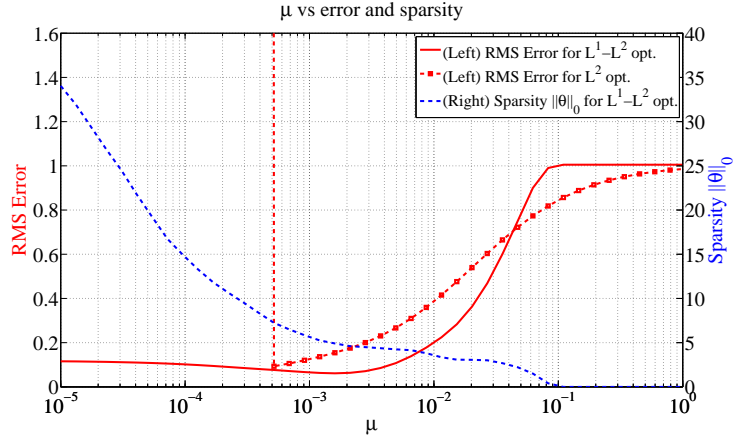}
}
\caption{Regularization parameter versus RMS tracking error and average sparsity}
\label{fig:mu}
\end{figure}

First, we compute the relation between the regularization parameter $\mu$
in (\ref{eq:J1})
and 
metrics for control performance
to be achieved by the optimal control vector
$\vc{\theta}[k]$.
We use two metrics: 
RMS (Root Mean Square) of the tracking error 
$e:=y - r$ and 
the average sparsity of control vector.
$\{\vc{\theta}[k]\}_{k = 0} ^{100}$. 
The RMS is defined as:
\[
 \rms{e} 
 := \sqrt{\frac{1}{T_{\text{f}}}\int_0^{T_{\text{f}}} |e(t)|^2 \dd t} 
 = \sqrt{\frac{1}{T_{\text{f}}}\sum_{k=0}^{100}\tnorm{y_k-r_k}_2^2}.
\]
The average sparsity is defined as:
$\|\vc{\theta}\|_0 := \sum_{k=0}^{100}\frac{\|\vc{\theta}[k]\|_0}{101}$.
Fig.~\ref{fig:mu} shows the performance as a function of the parameter $\mu$.
To compare the proposed method with a conventional one,
we consider the $L^2$-optimal control which minimizes 
$J_2(u_k) := \tnorm{y_k-r_k}_2^2 + \mu_2 \tnorm{u_k}_2^2$.
This cost function limits the energy (i.e., the $L^2$ norm) of the control $u_k$,
which has been widely used in control.
The optimal control vector, say $\vc{\theta}_2[k]$, is given by
\begin{equation}
 \vc{\theta}_2[k] = (\mu_2I + G^\top G)^{-1}G^\top (\vc{r}[k]-H\vc{x}[k]),
 \label{eq:l2opt}
\end{equation}
where we assume the control $u_k$ is in the subspace $\V_M$ defined in (\ref{eq:VM}).
The RMS error performance for the $L^2$-optimal control is also shown in Fig.~\ref{fig:mu}.
In this case, the feedback system becomes unstable for $\mu_2<0.0005$.
The sparsity of the optimal control vector is $\|\vc{\theta}_2[k]\|_0=101$
for all $k=0,1,\dots,101$.
That is, the $L^2$-optimal control does not produce any sparse vectors at all.

\begin{figure}[tb]
\centering
\mbox{
\subfigure[Proposed]{\includegraphics[width=0.4\linewidth]{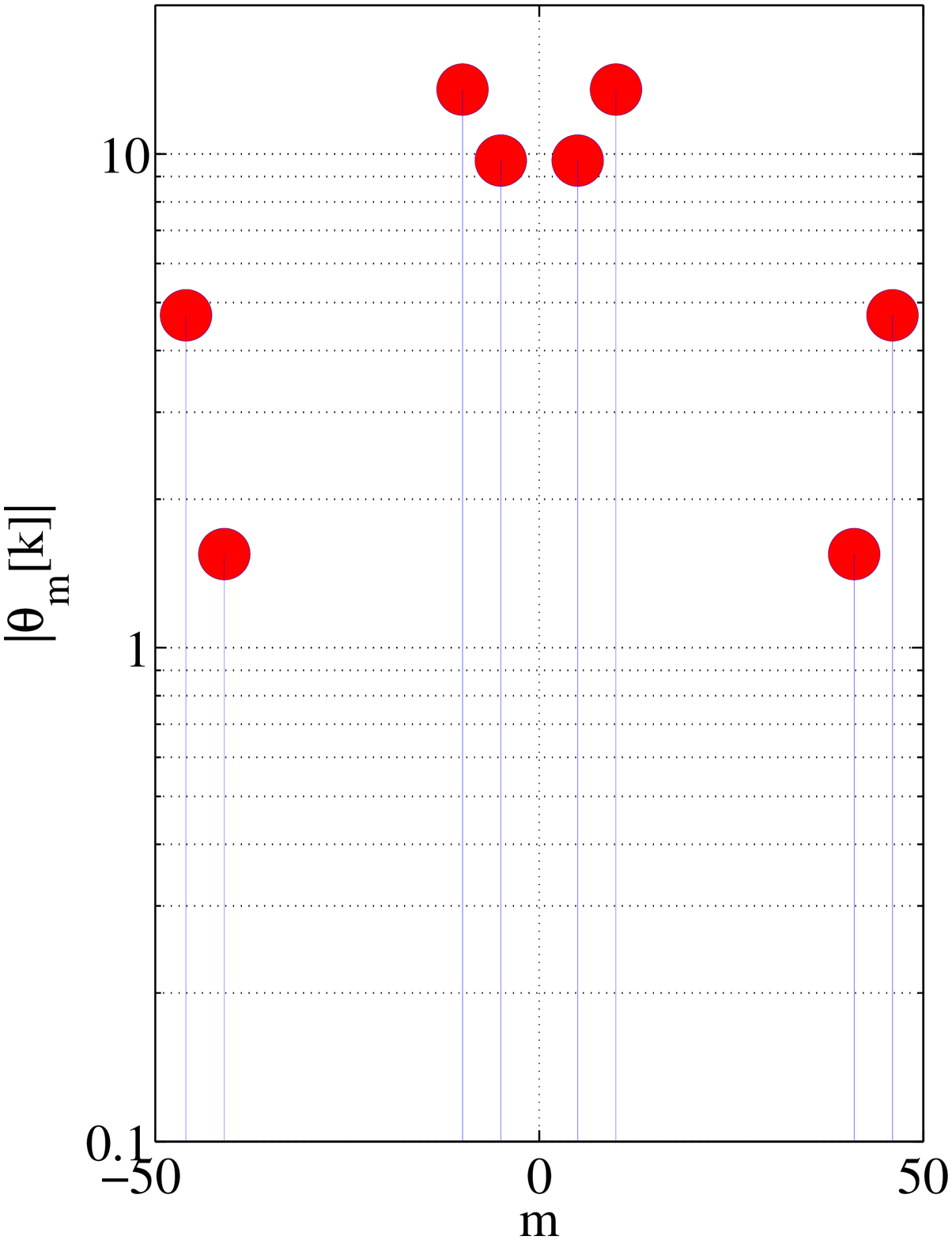}}
\subfigure[Conventional]{\includegraphics[width=0.4\linewidth]{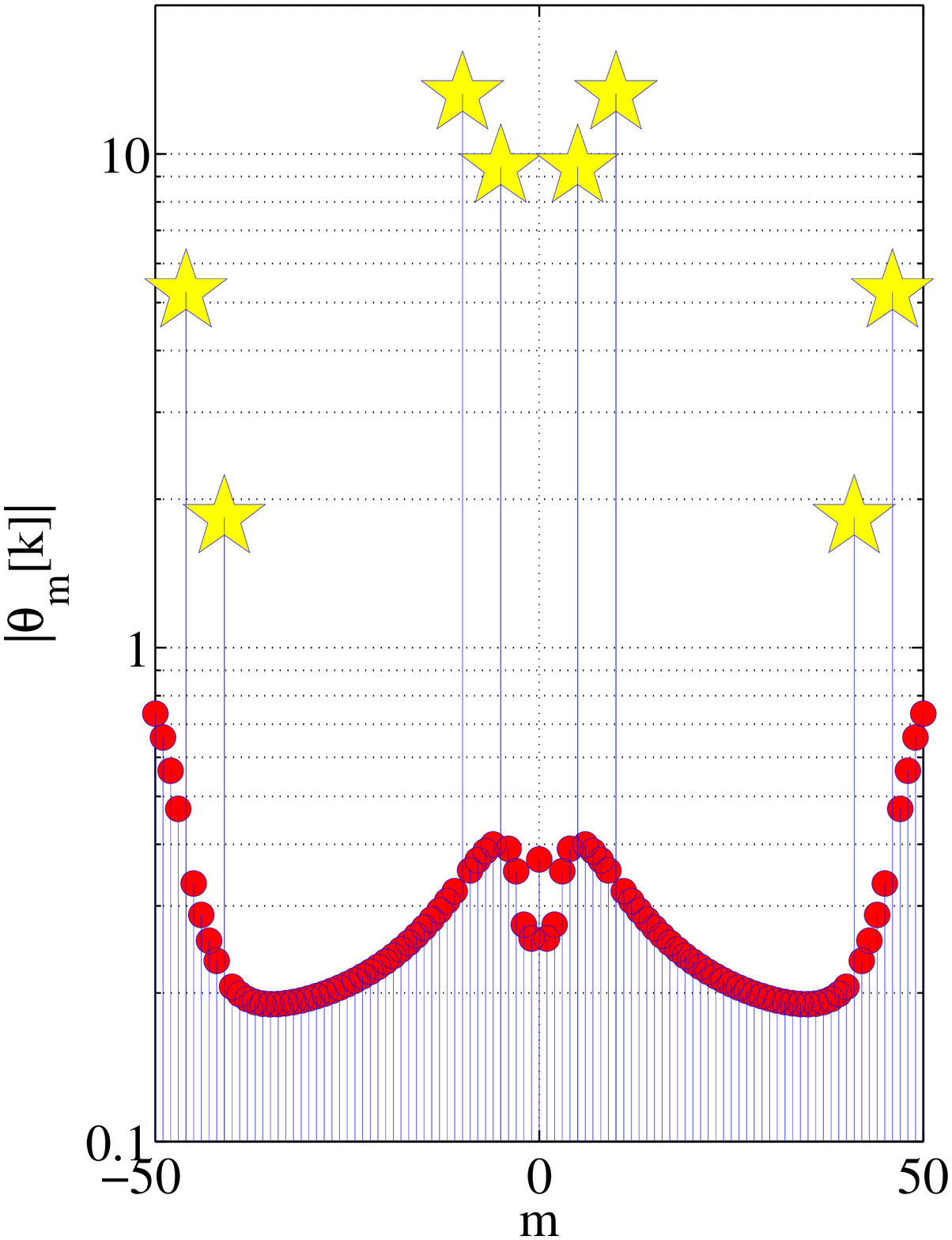}}
}
\caption{The absolute values of the nonzero coefficients of $\vc{\theta}[k]$
based on compressive sampling
with $\elll$ optimization
(left)and $\vc{\theta}_2$[k] 
based on $L^2$ optimal control (right) at $k=50$.
The stars in the right figure are the 8 elements of the truncated vector.
}
\label{fig:coef}
\end{figure}

\begin{figure}[tb]
\centering{
\includegraphics[width=0.88\linewidth]{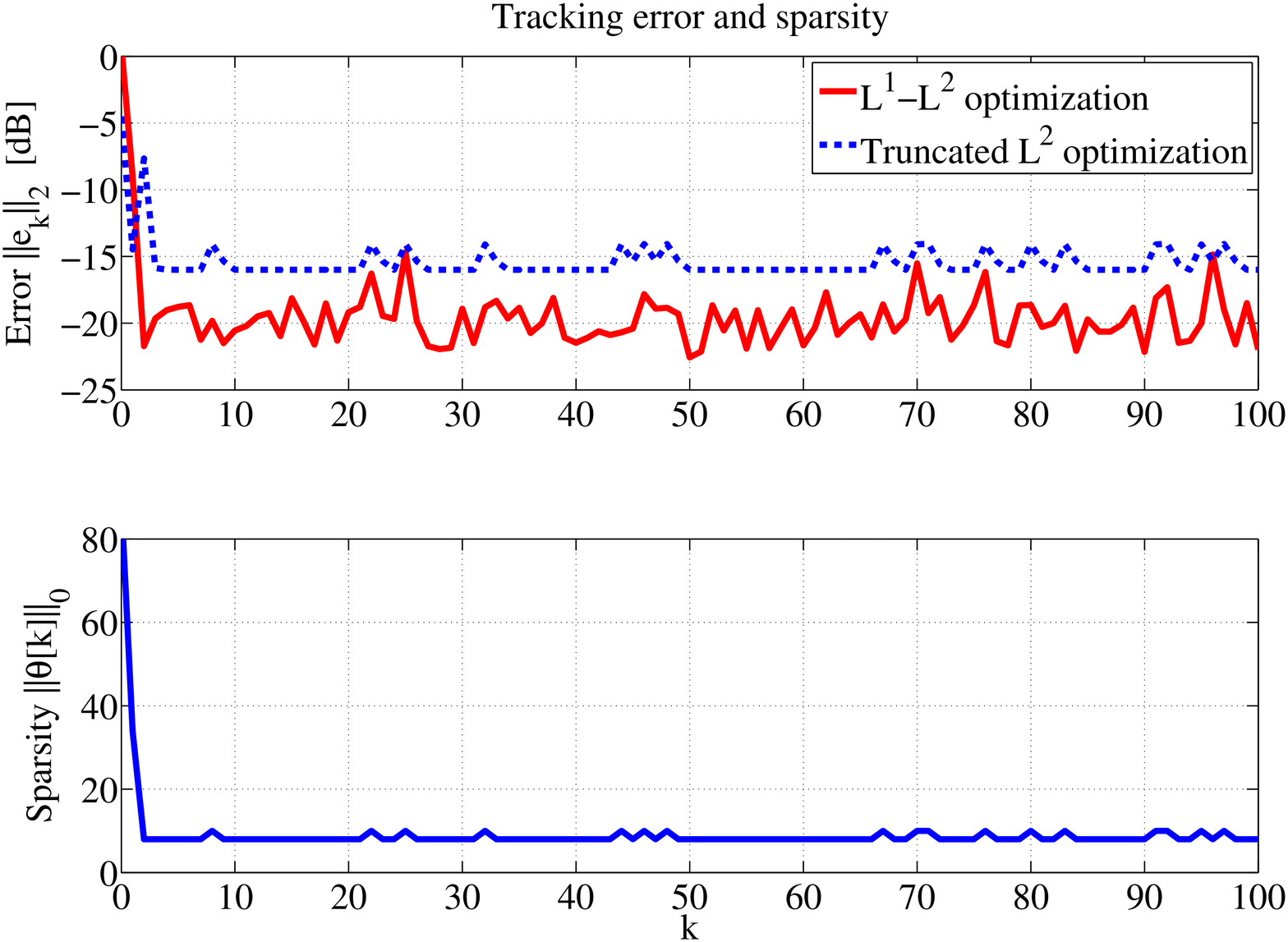}
}
\caption{Top figure: Tracking error $\|e_k\|_2$ in dB (solid: proposed, dash: conventional),
bottom figure: sparsity $\|\vc{\theta}[k]\|_0$}
\label{fig:comparison}
\end{figure}

Fig.~\ref{fig:mu} suggests that the optimal parameter is
$\mu\approx 0.002$ for $\elll$ optimization,
and
$\mu_2\approx 0.0005$ for $L^2$ optimization.
With these parameters, we simulate the feedback control.
Fig.~\ref{fig:coef}~(a) shows the absolute value of 
the nonzero elements in the $\elll$ optimal control vector $\vc{\theta}[k]$ at $k=50$.
We can see that the number of the nonzero elements is 8 out of 101 (the size of the vector),
and hence the vector is very sparse.
Then, the $L^2$ norm of the tracking error $e_k:=r_k-y_k$ on the $k$-th period
$\tnorm{e_k}_2$ ($k=0,1,\dots,100$)
is shown in the top figure in Fig.~\ref{fig:comparison}.
The sparsity history $\{\|\vc{\theta}[k]\|_0\}_{k=0}^{100}$ is shown
in the bottom figure in Fig.~\ref{fig:comparison}.

To compare the proposed method with conventional $L^2$-optimal control,
we compute the control vector $\vc{\theta}_2[k]$ by the formula (\ref{eq:l2opt})
with $\mu_2=0.0005$.
Fig.~\ref{fig:coef}~(b) shows the absolute value of the nonzero elements 
in the control vector $\vc{\theta}_2[k]$ at $k=50$.
We can see that all the elements in this vector are nonzero
(cf  Fig.~\ref{fig:coef}~(a)).
One may think that
the vector $\vc{\theta}_2[k]$ is {\em compressible}
since almost all the elements are nearly zero.
To see the difference,
we truncate the full vector $\vc{\theta}_2[k]$ 
by using the sparsity history in Fig.~\ref{fig:comparison}.

The stars in Fig.~\ref{fig:coef}~(b) are the 8 elements of the truncated vector.
The tracking error by the truncated vectors is shown in the top figure in Fig.\ref{fig:comparison}.
The proposed method shows the better performance than the truncated $L^2$-optimal control
with the same data size.
This shows the effectiveness of our method.

In an additional simulation study, we considered a step function for the reference,
that is,
$r_k(t) = r \in \re$, $k=0,1,2,\dots$.
This signal is also sparse in the space $\V_M$
and produces a sparse control vector
(we omit details due to space limitations).
\section{Conclusion}

We have studied the use of compressive sampling 
for feedback control systems with rate-limited communication channels.
Simulation studies indicate that the method proposed can effectively compress
the signals transmitted.
Control vectors are obtained via an $\elll$ optimization,
which is solved  by the FISTA algorithm.
Future work could include further investigation of bit-rate issues and the
study of closed loop stability.

\end{document}